\documentclass[prl,aps,twocolumn,longbibliography]{revtex4-2}

\usepackage{graphicx}
\usepackage{amsmath, amsxtra, amssymb, mathtools, isomath, nccmath, xspace, xcolor}
\usepackage[utf8]{inputenc}
\usepackage[T1]{fontenc} 
\usepackage{times} 

\usepackage{bm}	%
\usepackage{siunitx}
\usepackage{subfigure}
\usepackage{setspace, changepage}

\definecolor{myciteColor}{rgb}{0.39,0.7,0.89}
\usepackage[colorlinks=true,
citecolor=myciteColor,
linkcolor=myciteColor,urlcolor=myciteColor]{hyperref}

\newcommand{\kD}{k_{\rm D}}

\newcommand{\kF}{k_{\rm F}}
\newcommand{\kB}{k_{\rm B}}
\newcommand{\UD}{U_{\rm D}}

\newcommand{\omegaF}{\omega_{\rm F}}
\def\nobreakbefore{%
  \relax\ifvmode\else
    \ifhmode
      \ifdim\lastskip > 0pt\relax
        \unskip\nobreakspace
      \else 
        \nobreakspace
      \fi
    \fi
  \fi
}
\let\oldcite\cite
\renewcommand\cite{\nobreakbefore\oldcite}
\let\oldref\ref
\renewcommand\ref{\nobreakbefore\oldref}

\begin{document}
\title{
Universal equation of state for wave turbulence in a quantum gas
}

\author{
Lena~H.~Dogra$^1$, Gevorg~Martirosyan$^1$, Timon~A.~Hilker$^{1,2}$, Jake~A.~P.~Glidden$^1$,  Ji\v{r}\'{i}~Etrych$^1$, Alec~Cao$^1$, 
Christoph~Eigen$^1$,
Robert~P.~Smith$^3$, and Zoran~Hadzibabic$^1$
}

\date{\today}
\affiliation{$^1$ Cavendish Laboratory, University of Cambridge, J. J. Thomson Avenue, Cambridge CB3 0HE, United Kingdom\\
$^2$ Max-Planck-Institut f\"ur Quantenoptik, 85748 Garching, Germany\\
$^3$ Clarendon Laboratory, University of Oxford, Parks Road, Oxford OX1 3PU, United Kingdom}

\maketitle
{\bf
Boyle's 1662 observation that the volume of a gas is, at constant temperature, inversely proportional to pressure, offered a prototypical example of how an equation of state (EoS) can succinctly capture key properties of a many-particle system. Such relations are now cornerstones of equilibrium thermodynamics\cite{Landau:2013}. Extending thermodynamic concepts to far-from-equilibrium systems is of great interest in various contexts including glasses\cite{Cugliandolo:1997,Berthier:2011}, active matter\cite{Loi:2008, Takatori:2015,Ginot:2015,Fodor:2016}, and turbulence~\cite{Edwards:1969, Cardy:2008, Ruelle:2012, Picozzi:2014}, but is in general an open problem. Here, using a homogeneous ultracold atomic Bose gas\cite{Navon:2021}, we experimentally construct an EoS for a turbulent cascade of matter waves\cite{Zakharov:1992,
Nazarenko:2011}. Under continuous forcing at a large length scale and dissipation at a small one, the gas exhibits a non-thermal, but stationary state, which is characterised by a power-law momentum distribution\cite{Navon:2016} sustained by a scale-invariant momentum-space energy flux\cite{Navon:2019}. We establish the amplitude of the momentum distribution and the underlying energy flux as equilibrium-like state variables, related by an EoS that does not depend on the details of the energy injection or dissipation, or the history of the system. Moreover, we show that the equations of state for a wide range of interaction strengths and gas densities can be empirically scaled onto each other. This results in a universal dimensionless EoS that sets benchmarks for the theory and should also be relevant for other turbulent systems.
}

\begin{figure}[bp!]
		\includegraphics[width = \columnwidth]{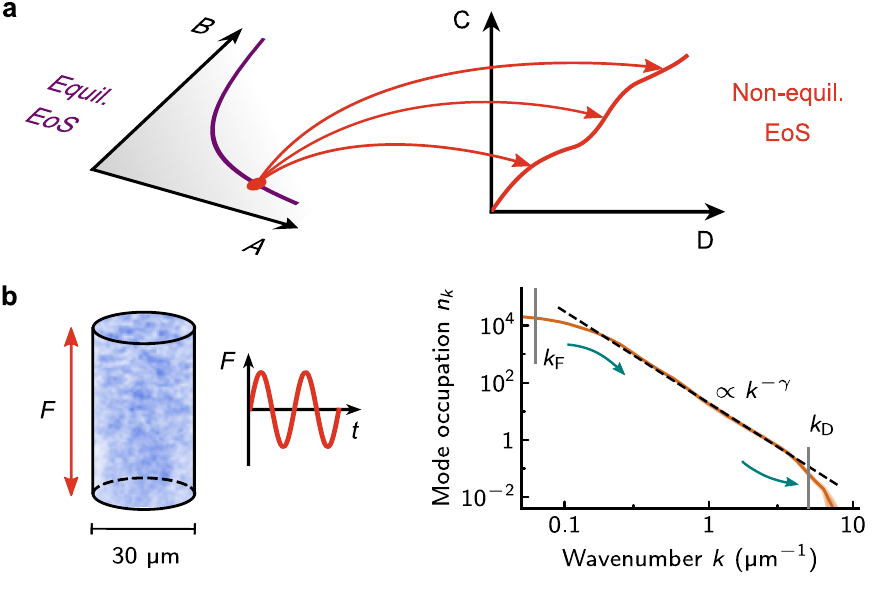}
\caption{
\textbf{Far-from-equilibrium equation of state and our experiment.}
{\bf a}, An EoS describes possible states of a macroscopic system by giving the relation between the state variables such as pressure or chemical potential.
Here, $A$ and $B$ are some generic equilibrium state variables, all equilibrium states lie in the $A-B$ plane, and out of each of them one can create (arrows) countless far-from-equilibrium ones. If the latter are stationary, they might still obey an EoS with new state variables $C$ and $D$. 
{\bf b}, Using an atomic Bose gas, we study a paradigmatic far-from-equilibrium stationary state, a turbulent cascade with matching energy injection at one length scale ($\kF^{-1}$) and dissipation at another ($\kD^{-1}$). Left: our gas is held in a cylindrical optical box (cartoon) and continuously driven on a large length scale by a time-periodic force $F$. Right: in steady state, the gas exhibits a highly non-thermal, but stationary, power-law momentum distribution $n_k\propto k^{-\gamma}$, with $\gamma = 3.2$ (Methods). 
\label{fig1}}
\end{figure}

The framework of thermodynamics provides an effective way to characterise equilibrium states of macroscopic systems without need for a detailed microscopic description. It can also be applied to near-equilibrium situations, such as linear transport, where the equilibrium state variables such as temperature are locally (in space and time) well defined.
A major ongoing challenge is to develop an equally effective framework for far-from-equilibrium systems. Such systems do not have all equilibrium variables defined even locally, but can nevertheless have well-defined stationary (albeit non-thermal) states, which in principle are amenable to thermodynamics-like treatments, including being describable by an equation of state (EoS) (Fig.~\ref{fig1}{\bf a}). Specifically, if quantities that describe fundamentally non-equilibrium phenomena, such as the energy dissipation rate, have values that are constant in time, they can take on the role of non-equilibrium state variables.

A turbulent cascade with matching energy injection (at one length scale) and dissipation (at a different one) is a paradigmatic stationary non-thermal state, sustained by a constant momentum-space energy flux that flows from the injection to the dissipation scale\cite{Richardson:1922}. 
From ocean waves\cite{Hwang:2000} to interplanetary plasmas\cite{Sorriso:2007} and financial markets\cite{Ghashghaie:1996}, such cascades generically result in power-law spectra of the various relevant quantities, with problem-dependent exponents.

For a given exponent, a cascade spectrum is fully defined by its amplitude. Famously, for hydrodynamic vortex turbulence in an incompressible fluid, dimensional analysis relates this amplitude to the magnitude of the underlying scale-invariant flux\cite{Kolmogorov:1941, Grant:1962, Sreenivasan:1995}; this amplitude-flux relation then serves as an equilibrium-like EoS. In general, dimensional analysis is insufficient, but for wave turbulence there are solvable approximate models that give more physical insight and also imply EoS-like amplitude-flux relations\cite{Zakharov:1992, Nazarenko:2011, Picozzi:2014}.

\begin{figure*}[ht!]
\centering
\includegraphics[width = \textwidth]{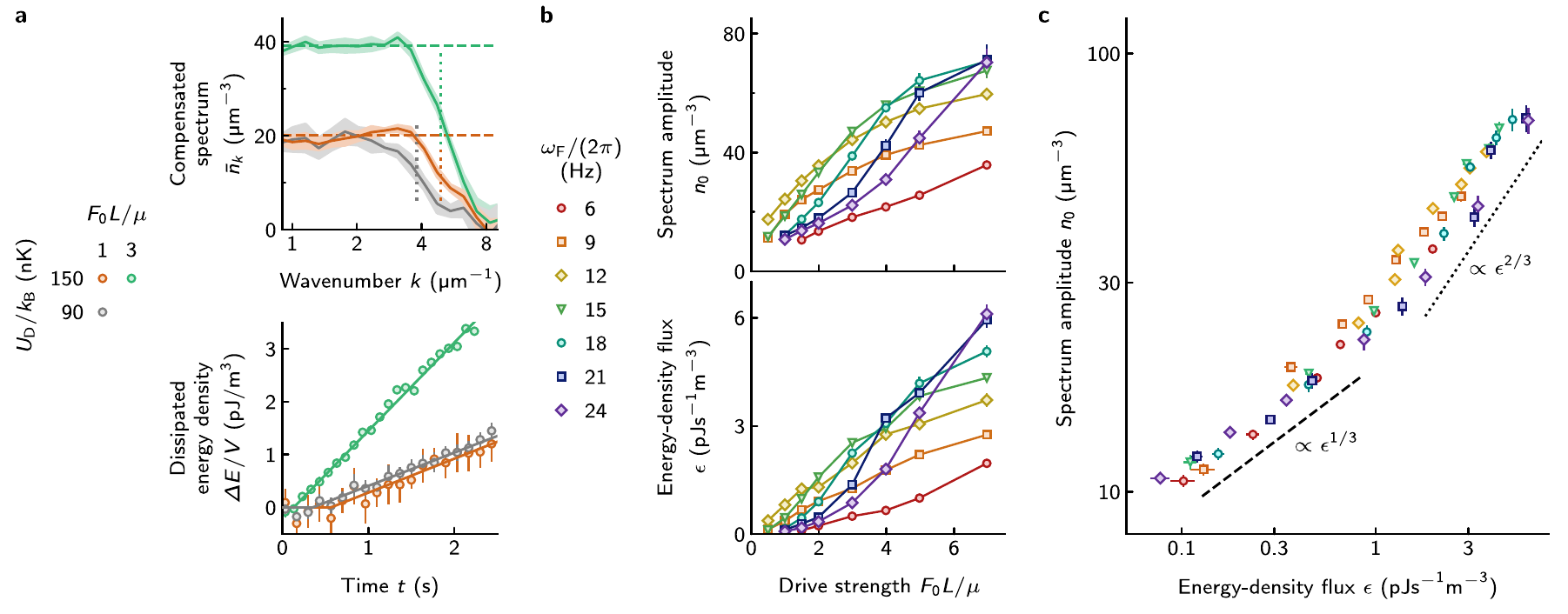}
\caption{
{\bf Constructing an equation of state for a turbulent quantum gas.}
Here we explore turbulent steady states in a gas with initial density $n = 5.7\,\si{\micro \meter^{-3}}$ and interaction strength $a=100\,a_0$; the chemical potential, $\mu \propto na$, is $\kB \times \SI{4.7}{\nano \kelvin}$, and the healing length is $\xi=\SI{1.2}{\micro \meter}$. The gas is driven by a spatially uniform force $F_0 \sin(\omegaF t)$, with variable $F_0$ and $\omegaF$. The trap depth $\UD$, which sets the dissipation scale $\kD \propto \sqrt{\UD}$, is $\kB \times 150$~nK, except for the grey data in {\bf a}, where it is $\kB \times 90$~nK.  
\textbf{a}, Examples of the steady-state momentum distributions $n_k$ and the corresponding energy fluxes $\epsilon$, for $\omegaF / (2\pi) = \SI{15}{Hz}$. Top: we plot the compensated spectra $\bar{n}_k = k^3 (k\xi)^{0.2} n_k$ (see text), so the spectrum amplitudes $n_0$ are seen in the plateaux indicated by the horizontal dashed lines; the vertical dotted lines indicate the $\kD$ values. Bottom: we plot the dissipated energy per unit volume, $\Delta E/V = \UD \Delta N /V$, where $\Delta N$ is the total number of atoms that have left the trap at $\kD$, and $V$ is the box volume, so $\epsilon$ (dissipation rates) are the slopes of the data; the solid lines show piece-wise linear fits. Both $n_0$ and $\epsilon$ increase with $F_0$, but are independent of $\kD$.
\textbf{b}, $n_0$ and $\epsilon$ as a function of $F_0$ for different $\omegaF$. The curves for different $\omegaF$ have different shapes and cross due to nonlinear effects of strong driving on the excitation resonance.
\textbf{c}, Eliminating the drive parameters $F_0$ and $\omegaF$, and directly relating $n_0$ to $\epsilon$, collapses all the data onto a single EoS-like curve.
The dashed and dotted lines are guides to the eye. The shading and error bars in {\bf a} reflect standard errors of measurement, while the error bars in {\bf b} and {\bf c} reflect standard fitting errors.
\label{fig2}}
\end{figure*}

We study wave cascades from small to large wavenumbers $k$ (large to small length scales) using a Bose gas held in a cylindrical optical-box trap\cite{Gaunt:2013, Eigen:2016} and driven on a system-size length scale by a spatially uniform time-periodic force $F$ (Fig.~\ref{fig1}{\bf b}, left)\cite{Navon:2016}. 
In steady state, the cascade is characterised by an isotropic momentum distribution (Fig.~\ref{fig1}{\bf b}, right): $n_k ({\bf k}) \propto k^{-\gamma}$, with $\gamma = 3.2(2)$ in agreement with the theory of weak wave turbulence (WWT)\cite{Zakharov:1992,Nazarenko:2011,Chantesana:2019, Zhu:2022}; here $n_k$ is the mode occupation, ${\bf k}$ is the wavevector and $k = |{\bf k}|$. 
In theory, $n_k = n_0\, k^{-3} f(k)$, where $n_0$ is the cascade amplitude and $f$ is a slowly varying dimensionless function, such that asymptotically $n_k \sim k^{-3}$ for $k \rightarrow \infty$, while in a finite (experimentally relevant) $k$-range $n_k$ is close to a power-law with an effective $\gamma$ slightly larger than $3$ 
(Methods). To experimentally extract $n_0$ from a finite $k$-range, we model $f$ by $(k\xi)^{-0.2}$, where $\xi \approx \SI{1}{\micro \meter}$ is the healing length 
(Methods). The cascade terminates by atoms leaving the trap at $\kD$ (the dissipation scale) set by the trap depth, and the rate at which they leave gives the steady-state energy-density flux $\epsilon$\cite{Navon:2019}. We explore the relationship between $n_0$ and $\epsilon$ for different $F$, $\kD$, box sizes, and microscopic gas parameters.

We start with an equilibrium Bose--Einstein condensate of $2 \times 10^5$ atoms of $^{39}$K in the lowest hyperfine ground state, held in a trap of length $L = \SI{50}{\micro \meter}$ and radius $R = \SI{15}{\micro \meter}$, so the gas density is $n = 5.7\,\si{\micro \meter^{-3}}$. Using the Feshbach resonance at $402.7$\,G~\cite{Etrych:2023} we set the $s$-wave scattering length $a$ to $100 \, a_0$, where $a_0$ is the Bohr radius, so the chemical potential is $\mu = 4\pi \hbar^2 na/m = \kB \times \SI{4.7}{\nano \kelvin}$ and $\xi= \hbar/\sqrt{2m\mu}=\SI{1.2}{\micro \meter}$; here $\hbar$ is the reduced Planck constant, $\kB$ the Boltzmann constant, and $m$ the $^{39}$K atom mass. The force $F = F_0 \sin(\omegaF t)$, created by a magnetic field gradient, primarily injects energy into the lowest phonon mode\cite{Navon:2016,Galka:2022}, at $\kF = \pi/L = \SI{0.06}{\micro \meter^{-1}} \ll 1/\xi$, so the natural scales for the drive strength $F_0$ and frequency $\omegaF$ are set, respectively, by $\mu/L$ and  $\sqrt{\mu/m}\, \kF \approx 2 \pi \times \SI{10}{Hz}$. The trap depth is $\UD = \kB \times \SI{150}{\nano \kelvin} \gg \mu$, so $\kD = \sqrt{2 m \UD}/\hbar = \SI{4.9}{\micro \meter^{-1}}$.

In Fig.~\ref{fig2}\textbf{a} we show measurements of the steady-state $n_k$ and $\epsilon$ for two drive strengths and fixed $\omegaF$. In the top panel we plot compensated spectra, $\bar{n}_k = k^3 (k\xi)^{0.2} n_k$, so the $n_0$ values are seen in the plateaux indicated by the dashed lines. In the bottom panel, the corresponding $\epsilon$ values are given by the slopes of the data, which are essentially constant over several seconds; initially the slope is zero (there is no dissipation) until the flux reaches $\kD$ and the steady state is established\cite{Navon:2019,Galka:2022}, while at long times (not shown) the condensate gets depleted. We also show, for $F_0 = \mu/L$, that neither $n_0$ nor $\epsilon$ change if we change $\kD$ by reducing $\UD/\kB$ to $90$~nK. 

In Fig.~\ref{fig2}\textbf{b} we present a systematic study of the cascade amplitudes and fluxes for different drive parameters $F_0$ and $\omegaF$. Both $n_0$ and $\epsilon$ monotonically increase with $F_0$ for any fixed $\omegaF$, but individually they depend in a complicated way on both $F_0$ and $\omegaF$, with the different-$\omegaF$ curves having different shapes and even crossing due to nonlinear effects of strong driving on the excitation resonance\cite{Navon:2016, Zhang:2021}.

However, as we show in Fig.~\ref{fig2}\textbf{c}, plotting $n_0$ versus $\epsilon$ reveals a unique EoS-like relation. All the data from Fig.~\ref{fig2}\textbf{b} collapse onto a single curve, showing that the steady-state $n_k$ depends only on the underlying flux and not on the details of its injection. Also note that $n_0$ is, for fixed $\kD$, proportional to the energy density, and hence pressure, so although we extracted it from the full microscopic $n_k$, it could in principle be a macroscopic observable.

In Fig.~\ref{fig2}{\bf c}, the low-$\epsilon$ data are consistent with the scaling $n_0 \propto \epsilon^{1/3}$ from perturbative WWT theory\cite{Zakharov:1992, Nazarenko:2011}; modelling a turbulent Bose gas by the classical-field Gross--Pitaevskii equation (GPE)\cite{Zakharov:1992, Nazarenko:2011, Navon:2016, Zhu:2022, Sano:2022} and assuming that the cascade transport is driven by four-wave mixing of incoherent waves, without any role played by the coherent condensate, gives an analytical prediction $\epsilon \propto  \hbar^{3} n_0^{3} a^{2}/m^2$. 
However, for large $\epsilon$ we observe significant departure from this scaling, suggesting qualitatively different behaviour.
Recently there has been a lot of interest in different regimes of turbulence in strongly driven condensates~\cite{Tsatsos:2016, Tsubota:2017, MiddletonSpencer:2022, Barenghi:2023}, but we are not aware of any theory that explains our results. Incidentally, our large-$\epsilon$ data are closer to $n_0 \propto \epsilon^{2/3}$ scaling, and the energy spectrum for the hydrodynamic vortex turbulence is\cite{Kolmogorov:1941} $E(k) \propto \epsilon^{2/3} k^{-5/3}$, but this similarity is likely fortuitous; for our system, numerical GPE simulations\cite{Navon:2016} show presence of some vortices, but our $E(k)\propto k^{4-\gamma}$ has a different $k$-dependence. 

\begin{figure}[t!]
\includegraphics[width = \columnwidth]{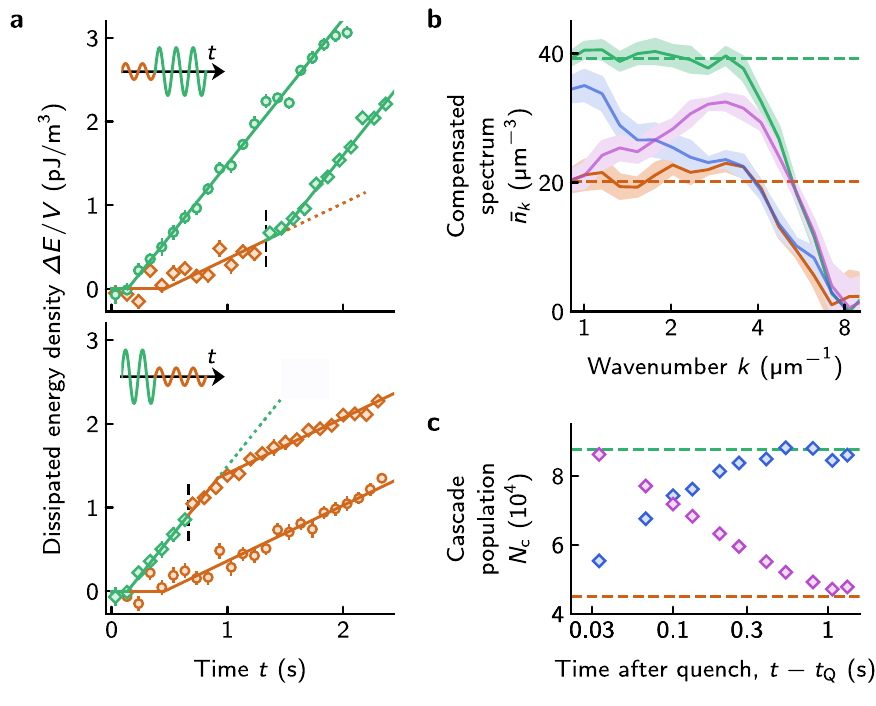}
\caption{
{\bf Switching between turbulent steady states.}
Quenches of the drive strength show that the steady-state $n_0$ and $\epsilon$, like equilibrium state variables, do not depend on the history of the system.
Here the parameters are the same as in Fig.~\ref{fig2}{\bf a} and we quench $F_0$ either from $\mu/L$ to $3\mu/L$ (`up quench') or vice versa (`down quench'). {\bf a}, The orange and green diamonds, respectively, correspond to the smaller and larger $F_0$, the black dashed lines indicate the quench times $t_{\rm Q}$, and the circles show reference data taken with constant $F_0$. For both up (top panel) and down (bottom panel) quench, the post-quench data is parallel to the reference one, showing that $\epsilon$ are the same. The small delays between $t_{\rm Q}$  and the changes in $\epsilon$ measured at $\kD$ reflect the need for the flux change to propagate through $k$-space. {\bf b}, The green and orange curves, respectively, show the steady-state compensated spectra after the up and down quenches, and the dashed lines show the reference $n_0$ values from Fig.~\ref{fig2}{\bf a}. We also show examples of transient $\bar{n}_k$, during the switching between the steady states, for the up (blue, $t - t_{\rm Q} = 0.03$~s) and down (purple, $t - t_{\rm Q} = 0.13$~s) quenches. {\bf c}, State-switching dynamics seen in the total number, $N_c$, of atoms in the cascade (Methods), for the up (blue) and down (purple) quenches; the dashed lines show steady-state values.  The shading in {\bf b} and the error bars  in {\bf a} and {\bf c} (often smaller than symbol sizes) show standard errors 
of measurement.
\label{fig3}}
\end{figure}

To further explore the analogy between equilibrium state variables and our $n_0$ and $\epsilon$, we study the response of a turbulent gas to dynamical changes in the driving force (Fig.~\ref{fig3}). In equilibrium, state variables have no memory of the history of the system. Here, we prepare one of the two steady states shown in Fig.~\ref{fig2}\textbf{a} (with $\UD/\kB = 150$~nK), then suddenly quench $F_0$, either from $\mu/L$ to $3\mu/L$ or vice versa, and show that the new steady state indeed has the same $\epsilon$ (Fig.~\ref{fig3}{\bf a}) and $n_0$ (Fig.~\ref{fig3}{\bf b}) as if $F_0$ had always been equal to its new value. 

We also briefly look at the state-switching dynamics, when the system `re-equilibrates' to the new (non-equilibrium) steady state. Following the quench of $F_0$, it takes a nonzero time for the flux change at $\kF$ to propagate to $\kD$\cite{Navon:2019, Galka:2022}, just like it takes a nonzero time to initially establish a turbulent steady state starting from equilibrium. In Fig.~\ref{fig3}{\bf b}, blue and purple curves illustrate, for the increased and decreased $F_0$ respectively, how the change in $n_k$ propagates from low to high $k$, with the local (in $k$-space) cascade population increasing for the blue curve and decreasing for the purple one; note that in the latter case some atoms return to low $k$. During the switching between steady states, $n_0$ is not defined, so to simply quantify the quench dynamics in Fig.~\ref{fig3}{\bf c} we show how the total cascade populations approach their new steady-state values. 

\begin{figure*}[t]
\centering
\includegraphics[width = \textwidth]{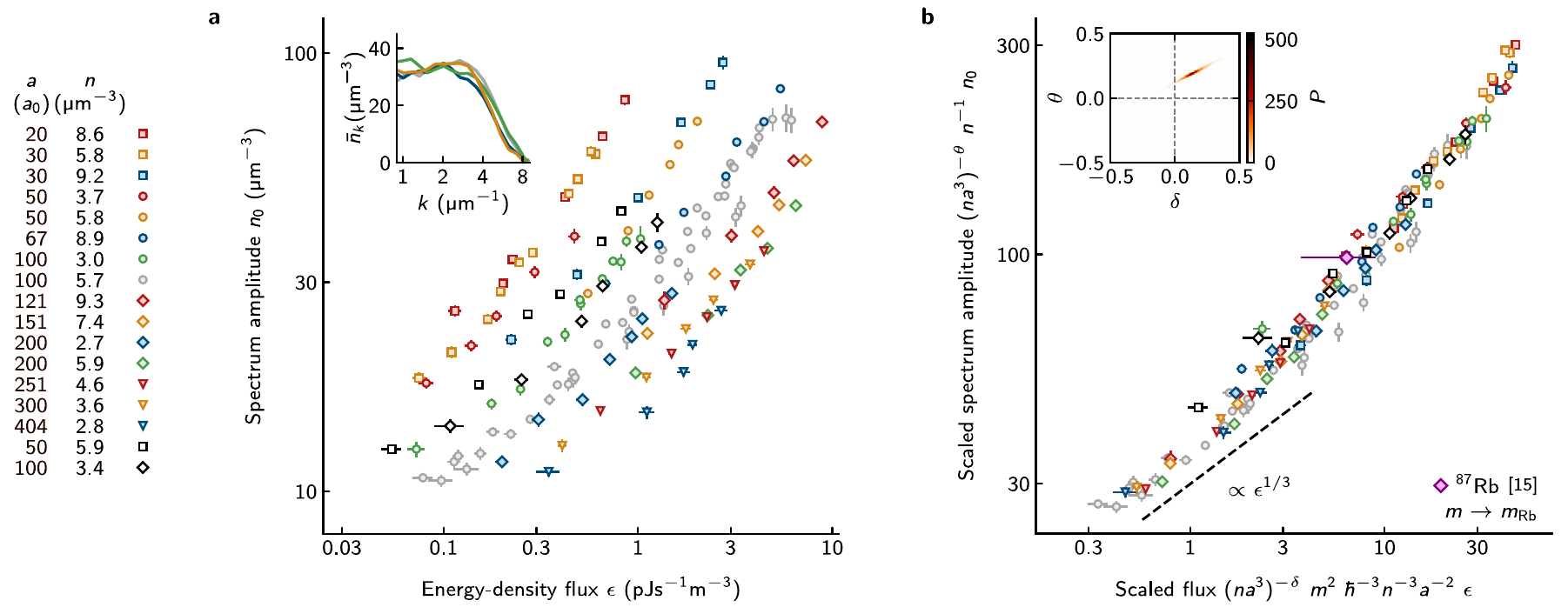}
\caption{
{\bf Universal equation of state.}
We generalise the measurements of $n_0(\epsilon)$ to different interaction strengths $a$ and gas densities $n$; all the data from Fig.~\ref{fig2}{\bf c} are here shown in grey. 
The black symbols indicate data sets for which the injection scale was also changed from $\kF = \SI{0.06}{\micro \meter^{-1}}$ by varying the box length (squares: $\kF = \SI{0.04}{\micro \meter^{-1}}$, diamonds: $\kF = \SI{0.08}{\micro \meter^{-1}}$). 
\textbf{a}, The EoS $n_0(\epsilon)$ is not universal for different $a$ and $n$. The inset shows four steady-state $\bar{n}_k$ with the same $n_0 \approx \SI{32}{\micro \meter^{-3}}$ but fluxes that vary between $0.28$ and $\SI{3.4}{\pico \joule \, \second^{-1} \meter^{-3}}$. 
\textbf{b}, We empirically collapse all the data onto a universal EoS by making both axes dimensionless and allowing additional scaling with the dimensionless parameter $na^3$ (see text). In the main panel $\delta = 0.13$ and $\theta=0.19$, while the inset shows the joint probability distribution $P$ for these scaling exponents (Methods). The single purple point shows data extracted from a $^{87}$Rb experiment\cite{Navon:2016}, with the appropriate mass scaling included ($m_{\rm Rb}$ denotes the $^{87}$Rb mass). The dashed line is a guide to the eye. The error bars show standard fitting errors. 
\label{fig4}}
\end{figure*}

Having established $n_0$ and $\epsilon$ as good state variables, in Fig.~\ref{fig4} we generalise our measurements of $n_0(\epsilon)$ to different interaction strengths and gas densities, {\it i.e.}, different initial equilibrium states (see Fig.~\ref{fig1}{\bf a}). For two data sets we also vary the box length, and hence $\kF = \pi/L$. In Fig.~\ref{fig4}\textbf{a} we show that the relationship between $n_0$ and $\epsilon$ is not unique for different $a$ and $n$; here, for the same $n_0$ the underlying flux varies by more than an order of magnitude.

However, we empirically find that all the data can be collapsed onto a universal EoS using simple scaling with $a$ and $n$ (Fig.~\ref{fig4}{\bf b}). Since the experimental EoS is not simply a power-law, even for single $(a,n)$ (Fig.~\ref{fig2}{\bf c}), it can be universal only in a completely dimensionless form, but this requirement does not lead to unique scaling predictions: $\epsilon$ can be scaled into dimensionless form by $\hbar^{3} n^{3} a^{2} m^{-2} (na^3)^{\delta}$ with any $\delta$, and $n_0$ can be scaled by $n (na^3)^{\theta}$ with any $\theta$. Using $\delta$ and $\theta$ as the only free parameters, we find good data collapse for $\delta = 0.13$ and $\theta = 0.19$ (Methods); note that here we do not presume the shape of the universal EoS, treat all the data points in Fig.~\ref{fig4}{\bf a} together as a single data set, and use $\delta$ and $\theta$ to simply minimise the data scatter. To further verify our results, we extract $n_0$ and $\epsilon$ from an experiment with a $^{87}$Rb gas\cite{Navon:2016} (Methods); this additional data point also agrees with our universal curve.

Even for low (scaled) $\epsilon$ these $a$ and $n$ scalings are not explained by the perturbative WWT theory, where the cascade dynamics depend only on $n_0$ and not on the total density $n$; for a direct comparison of the data with this theory see Extended Data Fig.~\ref{figS1}{\bf c}. 
The fact that the experimental EoS depends on the total $n$ implies that the presence of the condensate is also relevant for the cascade transport.
We also note that, even with the condensate included, no GPE-based model is compatible with our data, because in any such model the interaction strength enters only via the product $n \times a$, and $n_0/n$ can be a function of only $\epsilon/n$ and $na$.

Our experimentally constructed EoS provides both support and new challenges for non-equilibrium theories. The fact that a universal EoS for matter-wave turbulence exists at all provides a paradigmatic example of an equilibrium-like description of far-from-equilibrium matter, but the form of this EoS remains unexplained. Our experiments also demonstrate the possibility to study time-resolved transitions between non-thermal stationary states, which would be interesting to investigate further. In particular, an important question is how the transitions between two flux-carrying steady states relate to the turbulent relaxation and thermalization in isolated quantum systems\cite{Micha:2004, Berges:2008, Pruefer:2018, Erne:2018, Glidden:2020, GarciaOrozco:2022}. 
Another important open problem is whether intermittency\cite{Batchelor:1949, Newell:2001} plays a role in our turbulent gas.
Finally, our results could inspire and be relevant for future experiments with other quantum fluids; it would be interesting to study whether universal equations of state can be constructed for turbulence in systems such as Fermi gases, atomic superfluids with dipolar interactions, or the dissipative exciton-polariton condensates, and whether and how they differ from our EoS.

We thank Claudio Castelnovo, Jean Dalibard, Nishant Dogra, Kazuya Fujimoto, Maciej Ga\l{}ka, Giorgio Krstulovic, Nir Navon, Davide Proment, and Martin Zwierlein for helpful discussions. This work was supported by EPSRC [Grants No.~EP/N011759/1 and No.~EP/P009565/1], ERC (QBox and UniFlat) and STFC [Grant No.~ST/T006056/1]. T.~A.~H. acknowledges support from the EU Marie Sk\l{}odowska-Curie program [Grant No.~MSCA-IF- 2018 840081]. A. C. acknowledges support from the NSF Graduate Research Fellowship Program (Grant No. DGE2040434). C.~E. acknowledges support from Jesus College (Cambridge). R.~P.~S acknowledges support from the Royal Society. Z.~H. acknowledges support from the Royal Society Wolfson Fellowship.


%

\setcounter{figure}{0} 
\setcounter{equation}{0} 
\renewcommand\theequation{\arabic{equation}} 
\renewcommand\thefigure{\arabic{figure}} 
\renewcommand{\theHfigure}{A\arabic{figure}}
\renewcommand{\figurename}{EXTENDED DATA FIG.}

\subsection{Methods}

\begin{figure*}
    \centering
    \includegraphics[width = \textwidth]{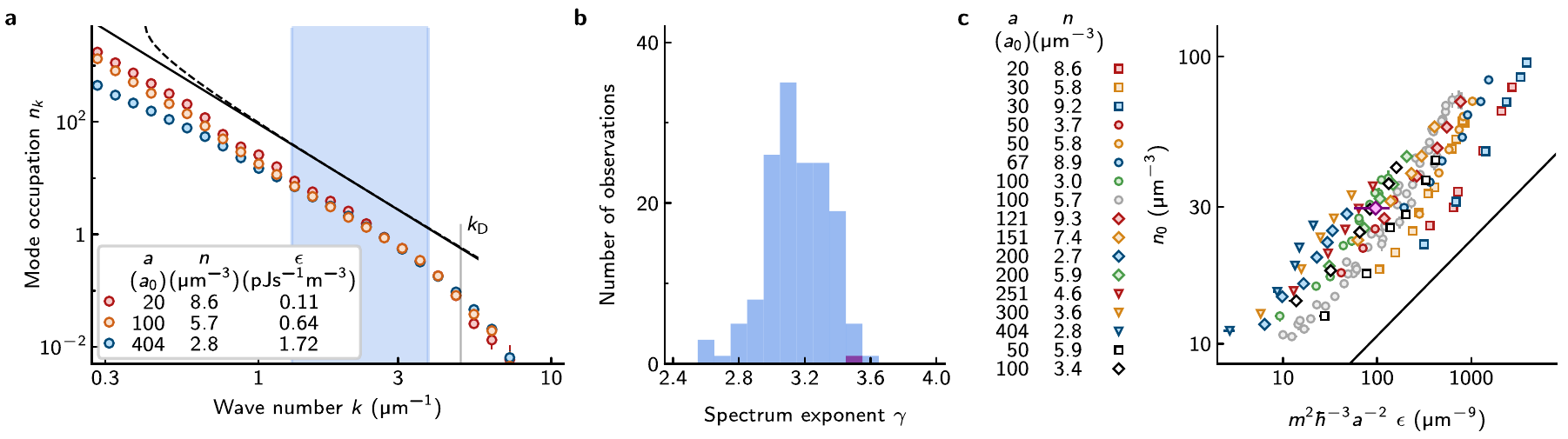}
\caption{
{\bf Steady-state momentum distributions and comparison with the perturbative WWT theory.} {\bf a}, Steady-state $n_k$ for three different combinations of $a$ and $n$, with fluxes $\epsilon$ chosen such that the cascade amplitudes are similar. The blue shading shows the $k$ range where we fit all our data. For the data shown here, $1/\xi =0.48 $ (red), $0.87$ (orange) and $\SI{1.22}{\micro\meter^{-1}}$ (blue). Fitted with $\gamma$ as a free parameter, these spectra give $\gamma = 3.3$ (red), $3.0$ (orange) and $3.1$ (blue). The solid line shows $n_0 k^{-3}(k\xi)^{-0.2}$ and the dashed line shows $n_0k^{-3}\ln(k/k_0)^{-1/3}$ with the same $n_0$ (chosen such as to offset the curves from the data for clarity), $\xi$ corresponding to the orange data, and $k_0 = \SI{0.4}{\micro \meter}^{-1}$. The error bars show standard errors of measurement.
{\bf b}, Histogram of extracted $\gamma$ for all spectra corresponding to the $153$ points in Fig.~\ref{fig4}, when fitting $n_k \propto k^{-\gamma}$ with
$\gamma$ as a free parameter. The purple bar indicates the Rb data point.
{\bf c}, Comparison of all the data shown in Fig.~\ref{fig4} with the perturbative WWT theory\cite{Zhu:2022} (solid line), without any free parameters. The error bars show standard fitting errors.
}
\label{figS1}
\end{figure*}

{\bf Steady-state momentum distributions.} 
The fact that the driven gas has reached its (quasi-)steady state is signalled by the onset of dissipation (atom loss; see Fig.~\ref{fig2}{\bf a}), and for consistency we always measure $n_k$ at the time when the atom number is reduced from its initial value by 15\%. We obtain $n_k$ from time-of-flight images, setting $a=0$ during the expansion, combining measurements for expansion times between $12$ and $78$~ms (with each measurement repeated $3$ - $6$ times), and reconstructing the three-dimensional distributions 
with the inverse Abel transformation\cite{Glidden:2020}. We normalise $n_k$ so that the total atom number is $N =  (2\pi)^{-3} V \int n_k \, 4\pi k^2 dk$.

As illustrated in Extended Data Fig.~\ref{figS1}{\bf a} for different experimental parameters, our $n_k$ are close to power-laws in a $k$-range $1/\xi \lesssim k\lesssim 0.8 \, \kD$; the examples shown here span the full range of $1/\xi$ values in Fig.~\ref{fig4}. To treat all data equally, we conservatively always fit $n_k$ between $k_{\rm min} = \SI{1.3}{\micro \meter^{-1}}$ and $k_{\rm max} = \SI{3.8}{\micro \meter^{-1}}$ (blue shading in Extended Data Fig.~\ref{figS1}{\bf a}), so $k_{\rm min} > 1/\xi$ is always satisfied. Similarly, the cascade population $N_c$ in Fig.~\ref{fig3} is defined as the atom number at $k > k_{\rm min}$.

Extended Data Fig.~\ref{figS1}{\bf b} shows the histogram of the fitted $\gamma$ values for all the spectra corresponding to the 153 data points in Fig.~\ref{fig4}; we find a mean $\gamma=3.2$ with a standard deviation of $0.2$. Various analytical and numerical calculations\cite{Zakharov:1992, Zhu:2022, Nazarenko:2011, Chantesana:2019} give that in our finite $k$ range one expects an effective $\gamma - 3 \lesssim 0.5$. Specifically, the analytical WWT prediction\cite{Zakharov:1992, Zhu:2022} for $n_0(\epsilon)$ is based on $n_k = n_0 k^{-3} f(k)$ with $f(k) = \ln(k/k_0)^{-1/3}$. In this calculation, particles are injected isotropically at $k_0$ and the results formally hold for $k\gg k_0$. We inject energy (anisotropically) at a very low $\kF$, in the phonon regime, but the onset of the isotropic cascade should still be at $k\sim 1/\xi$\cite{Galka:2022,Sano:2022}. Empirically, varying $k_0$ between $\SI{0.06}{\micro \meter}^{-1}$ (our $\kF$) and $\SI{0.6}{\micro \meter}^{-1}$ (just a factor of two smaller than $k_{\rm min}$), the analytical curves are, between $k_{\rm min}$ and $k_{\rm max}$, always fitted well by power laws and give effective $\gamma$ between $3.1$ and $3.3$;
our mean $\gamma = 3.2$ is reproduced by setting $k_0 = \SI{0.4}{\micro \meter^{-1}}$ (see Extended Data Fig.~\ref{figS1}{\bf a}).

To consistently extract the cascade amplitudes $n_0$ with dimensions of $\si{\micro \meter}^{-3}$, we refit all the data with fixed $\gamma=3.2$ and define $n_0$ via $n_k = n_0 \, k^{-3} (k\xi)^{-0.2}$, {\it i.e.}, we model $f(k)$ by $(k\xi)^{-0.2}$ in our fitting range. We use $\xi$ as the natural scale for making $f$ dimensionless, but using a constant $\bar{\xi}= \SI{1.3}{\micro \meter}$ (the geometric mean of our $\xi$ range) would change the $n_0$ values by only $\pm 10\%$ and not alter any conclusions. Modelling $f = (Ak\xi)^{-0.2}$ with a constant dimensionless $A \neq 1$ would be equally valid, simply rescaling $n_0$ by a constant and not affecting the EoS shape, but $A=1$ is both the simplest choice and makes our heuristic $f$ almost equal (within $\lesssim 10\%$ in our fitting range) to the analytical WWT one with $k_0 = \SI{0.4}{\micro \meter^{-1}}$ (see Extended Data Fig.~\ref{figS1}{\bf a}), which allows a fair comparison of our $n_0$ with this theory (Extended Data Fig.~\ref{figS1}{\bf c}).

{\bf Comparison with the perturbative WWT theory.}
Extended Data Fig.~\ref{figS1}{\bf c} shows the comparison of all our experimental $n_0$ values for different $a$ and $n$ (Fig.~\ref{fig4}) with the most recent perturbative WWT calculation\cite{Zhu:2022} (dashed line), without any free parameters. The theoretical and experimental $n_0$ agree within a factor of 3, but the theoretical scaling $n_0 \propto \left(m^2 \hbar^{-3} a^{-2} \epsilon \right)^{1/3}$, which does not depend on $n$, does not collapse the data onto a single curve.

{\bf The universal EoS.}
To find the optimal scaling exponents in Fig.~\ref{fig4}, for any given $(\delta, \theta)$ we quantify the collapse of the scaled data using the reduced $\chi^2$ of a simple piece-wise power-law fit (allowing for four $x$-axis regions with different power laws) to all the data points, without distinguishing different $a$ and $n$. For $\delta=\theta=0$ we get $\chi^2=47$, while for the optimal $\delta = 0.13$ and $\theta = 0.19$ we get $\chi^2 = 3.8$. We get essentially the same $\chi^2 = 3.5$ for the single-$(a,n)$ data series shown in Fig.~\ref{fig2}{\bf c}, which suggests that with optimal $\delta$ and $\theta$ all the dependence on $a$ and $n$ has been scaled out. The fact that these $\chi^2$ are larger than $1$ suggests that the data scatter is not purely statistical, but also comes from systematic errors. These could arise, for example, due to the small residual inhomogeneity of gases trapped in optical boxes\cite{Navon:2021}. For completeness, note that for the WWT scaling in Extended Data Fig.~\ref{figS1}{\bf c} we get $\chi^2 = 57$.

To construct the joint probability distribution $P(\delta, \theta)$ (inset of Fig.~\ref{fig4}{\bf b}) we randomly select $1/3$ of the data, apply the same optimisation procedure, and repeat this $10^4$ times. Treating $\delta$ and $\theta$ as independent variables with Gaussian distributions gives standard deviations $\sigma_{\delta} = 0.08$ and $\sigma_{\theta} = 0.04$. However, the errors in the two exponents are correlated, as seen from the shape of $P(\delta, \theta)$, and the peak probability density, $P(0.13, 0.19) \approx 500$, is ten times larger than the Gaussian result $1/(2\pi \sigma_{\delta} \sigma_{\theta}) \approx 50$.

{\bf Rb data point in Fig.~4.}
The Rb point further illustrates the universality of the EoS because it is a measurement with an atom of a different mass, performed with a different experimental apparatus.
We extracted this data point from Ref.\cite{Navon:2016}; we obtained $n_0$ from the spectrum shown in Fig.~3{\bf a} of that paper (applying the inverse Abel transform and fitting it with $\gamma = 3.2$) and deduced $\epsilon$ from the populations shown in the inset of Fig.~3{\bf b} in that paper. As for our data, for scaling $n_0$ and $\epsilon$ in Fig.~4{\bf b} we use the initial $n$.

\vspace{2em}


{\bf Data availability}\quad
The data that support the findings of this study are available in the Apollo repository (\href{https://doi.org/10.17863/CAM.96408}{https://doi.org/10.17863/CAM.96408}). Any additional information is available from the corresponding authors upon reasonable request.

{\bf Author contributions}\quad L.~H.~D. led the data collection and analysis, with most significant contributions from G.~M. and T.~A.~H. All authors (L.~H.~D., G.~M., T.~A.~H., J.~A.~P.~G., J.~E., A.~C., C.~E., R.~P.~S., and Z.~H.) contributed significantly to the experimental setup, the interpretation of the results and the production of the manuscript. Z.~H. supervised the project.

{\bf Competing interests}\quad The authors declare no competing interests.

{\bf Correspondence and requests for materials} should be addressed to L.~H.~D. (lhb31@cam.ac.uk), C.~E.~(ce330@cam.ac.uk), or Z.~H.~(zh10001@cam.ac.uk).

{\bf Reprints and permissions information} is available at www.nature.com/reprints.

\end{document}